\documentclass[amsmath,amssymb,aps,floats,prl,twocolumn,showpacs,unsortedaddress,superscriptaddress]{revtex4}

\usepackage{graphicx}
\usepackage{amssymb}
\usepackage{epstopdf}
\usepackage{color}
\usepackage[normalem]{ulem}

\newcommand{\be}{\begin{eqnarray}}
\newcommand{\ee}{\end{eqnarray}}
\newcommand{\bfig}{\begin{figure}}
\newcommand{\efig}{\end{figure}}

\newcommand{\Mtau}{M_{\tau}}
\newcommand{\Tcw}{T_{\mathrm{CW}}}
\newcommand{\rucl} {RuCl$_3$}
\newcommand{\ka} {\kappa}
\newcommand{\Hmin}{H_{\mathrm{min}}}

\begin{document}

\title{Anomalous thermal conductivity and magnetic torque response in the 
honeycomb magnet $\alpha$-RuCl$_3$ }

\author{Ian A. Leahy}
\affiliation{Department of Physics, University of Colorado, Boulder, CO 80309, 
USA}%
\author{Christopher A. Pocs}
\affiliation{Department of Physics, University of Colorado, Boulder, CO 80309, 
USA}%
\author{Peter E. Siegfried }
\affiliation{Department of Physics, University of Colorado, Boulder, CO 80309, 
USA}%
\author{David Graf }
\affiliation{National High Magnetic Field Laboratory, Tallahassee, FL 32310, 
USA}%
\author{S.-H. Do}
\affiliation{Department of Physics, Chung-Ang University,  Seoul, 790-784, 
South Korea}%
\author{Kwang-Yong Choi}
\affiliation{Department of Physics, Chung-Ang University,  Seoul, 790-784, 
South Korea}%
\author{B. Normand}
\affiliation{Laboratory for Neutron Scattering and Imaging, Paul Scherrer
Institute, CH-5232 Villigen PSI, Switzerland}%
\author{Minhyea Lee}
\affiliation{Department of Physics, University of Colorado, Boulder, CO 80309, 
USA}%

\date{\today}

\begin{abstract}
We report on the unusual behavior of the in-plane thermal conductivity 
($\ka$) and torque ($\tau$) response in the Kitaev-Heisenberg material 
$\alpha$-\rucl. $\ka$ shows a striking enhancement with linear growth 
beyond $H = 7$ T, where magnetic order disappears, while $\tau$ for both 
of the in-plane symmetry directions shows an anomaly at the same field. 
The temperature- and field-dependence of $\ka$ are far more complex than 
conventional phonon and magnon contributions, and require us to invoke 
the presence of unconventional spin excitations whose properties are 
characteristic of a field-induced spin-liquid phase related to the 
enigmatic physics of the Kitaev model in an applied magnetic field. 
\end{abstract}

\maketitle

Low-dimensional spin systems display a multitude of quantum phenomena, 
providing an excellent forum for the exploration of unconventional ground 
states and their exotic excitations. The Kitaev model \cite{Kitaev2006} has 
attracted particular attention, both theoretically and experimentally, 
because it possesses an exactly solvable quantum spin-liquid (QSL) ground 
state and has possible realizations in a number of candidate materials 
\cite{Jackeli2009,Chaloupka2010,Singh2012,Plumb2014}. Thermal transport 
measurements have proven to be a powerful tool for elucidating the itinerant 
nature of QSLs \cite{YamashitaM2008,YamashitaM2010}, as a result of their high 
sensitivity to the low-energy excitation spectrum, and in fact studies of 
low-dimensional insulating quantum magnets have revealed very significant 
contributions to heat conduction from unconventional spin excitations 
\cite{Ando1998,Sologubenko2000,Hofmann2001, Sologubenko2001, Sales2002,
Jin2003,Li2005,HessReview2007,Sun2009, Chen2011,Mohan2014}.  

Magnetic insulators containing $4d$ and $5d$ elements combine electronic 
correlation effects with strong spin-orbit coupling (SOC) to generate 
complex magnetic interactions. In the Kitaev model, nearest-neighbor 
spin-$\frac{1}{2}$ entities on a two-dimensional (2D) honeycomb lattice 
interact through a bond-dependent Ising-type coupling of different spin 
components, whose strong frustration leads to a QSL ground state with 
emergent gapless and gapped Majorana-fermion excitations \cite{Kitaev2006}. 
The physical realization of this uniquely anisotropic interaction requires 
strong SOC, which creates effective $j_{\mathrm{eff}} = \frac{1}{2}$ moments 
with Kitaev-type coupling in the honeycomb iridate compounds A$_2$IrO$_3$ 
(A = Na or Li) \cite{Choi2012, Chun2015}. Despite its weaker SOC, the $4d$ 
honeycomb material $\alpha$-\rucl~contains similar spin-orbit-entangled 
moments, and thus has emerged as another candidate system for Kitaev-related 
physics \cite{Plumb2014, Sears2015, Rau2014, Kim2015}. 

In this Letter, we present in-plane thermal conductivity ($\ka$) and magnetic 
torque ($\tau$) studies of single-crystal $\alpha$-\rucl~samples. Below the 
magnetic ordering temperature, $T_C$, a pronounced minimum of $\ka$ and an 
accompanying torque anomaly at $H = \Hmin \simeq 7$ T occur due to a 
field-induced phase transition from the ``zig-zag" ordered state 
\cite{Sears2015, Johnson2015} to a spin-disordered phase. The abrupt 
and linear rise of the low-$T$ $\ka$ at $H > \Hmin$ indicates that this 
field-induced spin liquid (FISL) contains a massless excitation with 
Dirac-type dispersion, while the strong renormalization of the phonon 
contributon at all temperatures suggests a broad band of unconventional 
medium-energy excitations. These results serve to fingerprint the possible 
Kitaev physics of the FISL in $\alpha$-\rucl.

Single crystals of \rucl~were synthesized by vacuum sublimation 
\cite{Park2016}, as described in Sec.~SI of the Supplementary Material 
(SM) \cite{SuppInfo}. $\ka$ measurements were performed with a 
one-heater, two-thermometer configuration in a $^3$He refrigerator and 
external magnetic fields up to 14 T. Cernox and RuO$_x$ resistors were used 
as thermometers for the respective temperature ranges $T > 2$ K and $0.3 < T 
\le 15$ K, and were calibrated both separately and in-situ under the applied 
field. Both the thermal current ($-\nabla T$) and the field were oriented in 
the crystalline $ab$-plane, with $H$ applied either parallel or perpendicular 
to $\nabla T$. We found little difference in $\ka$ for the two orientations, 
and all results shown below were measured in the $\nabla T \parallel H 
\parallel ab$ geometry, other than Fig.~\ref{kappaT}(b), where $\nabla T 
\perp H$. $\tau$ was determined from capacitance measurements between the 
ground plane and a BeCu cantilever. Its angular dependence was measured in 
two geometries, one in which $H$ was rotated within the $ab$-plane ($\phi$ 
rotation) and one with $H$ rotated out of plane ($\theta$ rotation).

\begin{figure}[t]
\begin{center}
\includegraphics[width=0.95\linewidth]{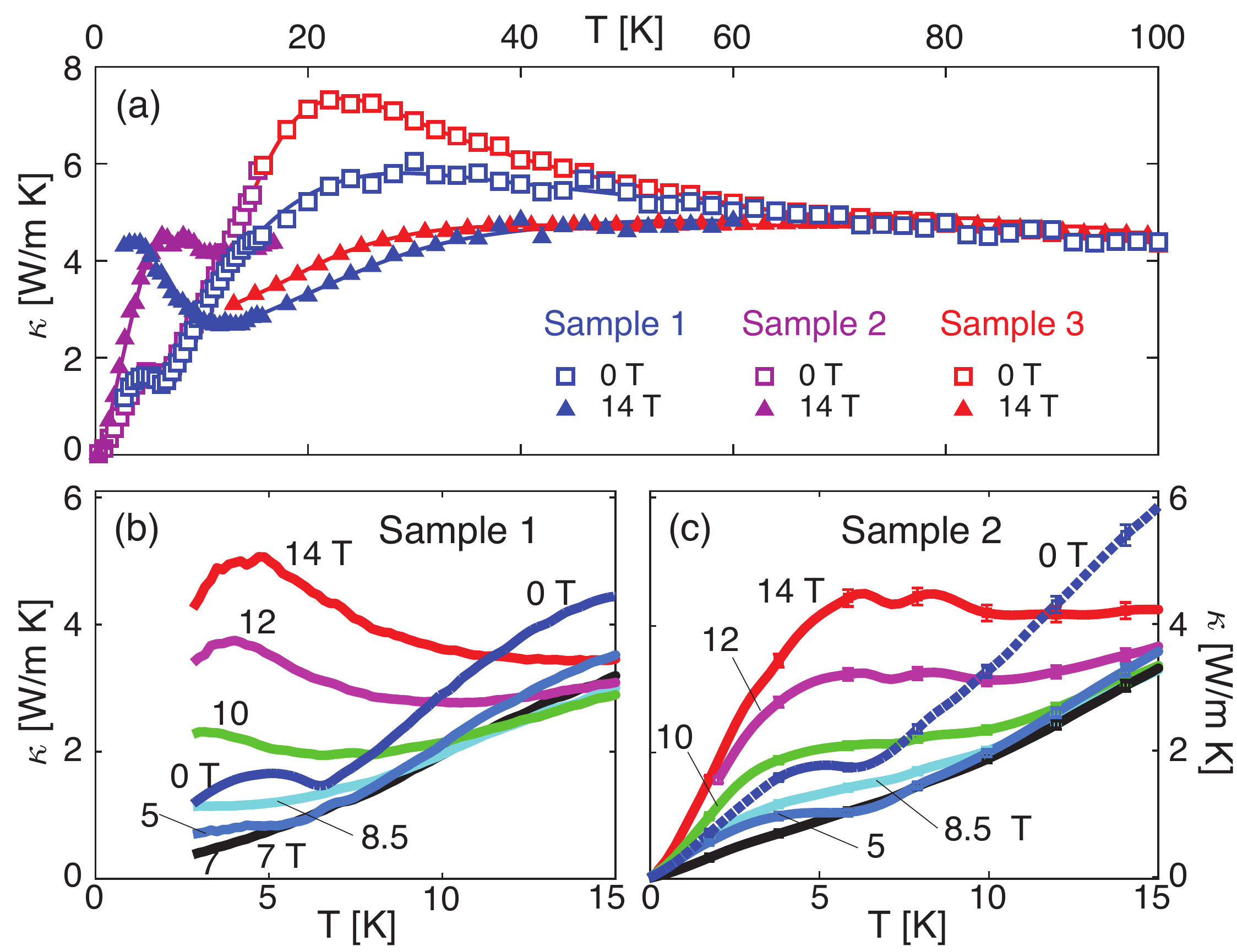}
\caption {(a) In-plane thermal conductivity, $\ka(T)$, shown up to 100 K 
for Samples 1, 2, and 3 at $\mu_0 H = 0$ T (squares) and 14 T (triangles). 
Solid lines are a guide to the eye. (b) Low-temperature detail of $\ka(T)$ 
for a range of $H$ values, shown for Sample 1. (c) $\ka(T)$ at low $T$ for 
Sample 2.}
\label{kappaT}
\end{center}
\efig

Figure \ref{kappaT}(a) shows $\ka(T)$, in fields $\mu_0H = 0$ and 14 T, for 
three $\alpha$-\rucl~ samples. Qualitatively, the dependence of $\ka$ on both 
$T$ and $H$ is the same in each case, and we focus on these general features. 
Quantitatively, our samples show differences in peak heights and widths, 
which we relate to their age and defect content in Sec.~SI of the SM 
\cite{SuppInfo}. On cooling at zero field (ZF), $\ka_0 (T) = \ka(T,H = 0)$ 
has a broad peak near 25 K and decreases down to the magnetic ordering 
temperature, $T_C \simeq 6.3$ K, which is identified both from the upturn in 
$\ka$ and from the magnetic susceptibility (data not shown). This value of 
$T_C$ is identified clearly in all our crystals, testifying to their high 
as-grown quality, with no contamination from structures of different layer 
stackings \cite{Cao2016}. For $T < T_C$, $\ka_0(T)$ shows a weak maximum 
before decreasing to zero. $\ka(T,14$ T) differs dramatically from $\ka_0(T)$ 
at all temperatures below 60 K. Its peak at intermediate $T$ is suppressed, 
broader, and lies at a higher temperature, whereas below $T \simeq 12$ K it 
has a strong low-$T$ peak that is completely absent from $\ka_0(T)$. 

Focusing on this low-$T$ regime, Figs.~\ref{kappaT}(b) and \ref{kappaT}(c) 
show $\ka(T)$ at constant fields $H = 0$, 5, 7, 8.5, 10, 12, and 14 T. 
Because the ordered state has a large magnon gap \cite{Banerjee2015}, the 
weak low-$T$, low-$H$ feature is in fact an enhanced phonon contribution. This 
is suppressed by increasing field, and the minimum marking $T_C$ is visible up 
to $H = 5$ T. At $H = \Hmin \simeq 7$ T, both the phonon enhancement and the 
minimum disappear. Further increase of $H$ causes the appearance of the 
low-$T$ peak, whose height grows linearly with $H - \Hmin$, leading to 
rounded maxima around 5 K at 14 T. We have collected detailed low-$T$ data 
($0.3 < T < 3$ K) at $H > \Hmin$ for Sample 2 [Fig.~\ref{kappaT}(c)] and find 
that these do not display an activated form; the alternative of a power-law 
form demonstrates clearly that this feature is the contribution of a gapless 
excitation. 

\begin{figure}[t]
\begin{center}
\includegraphics[width=1\linewidth]{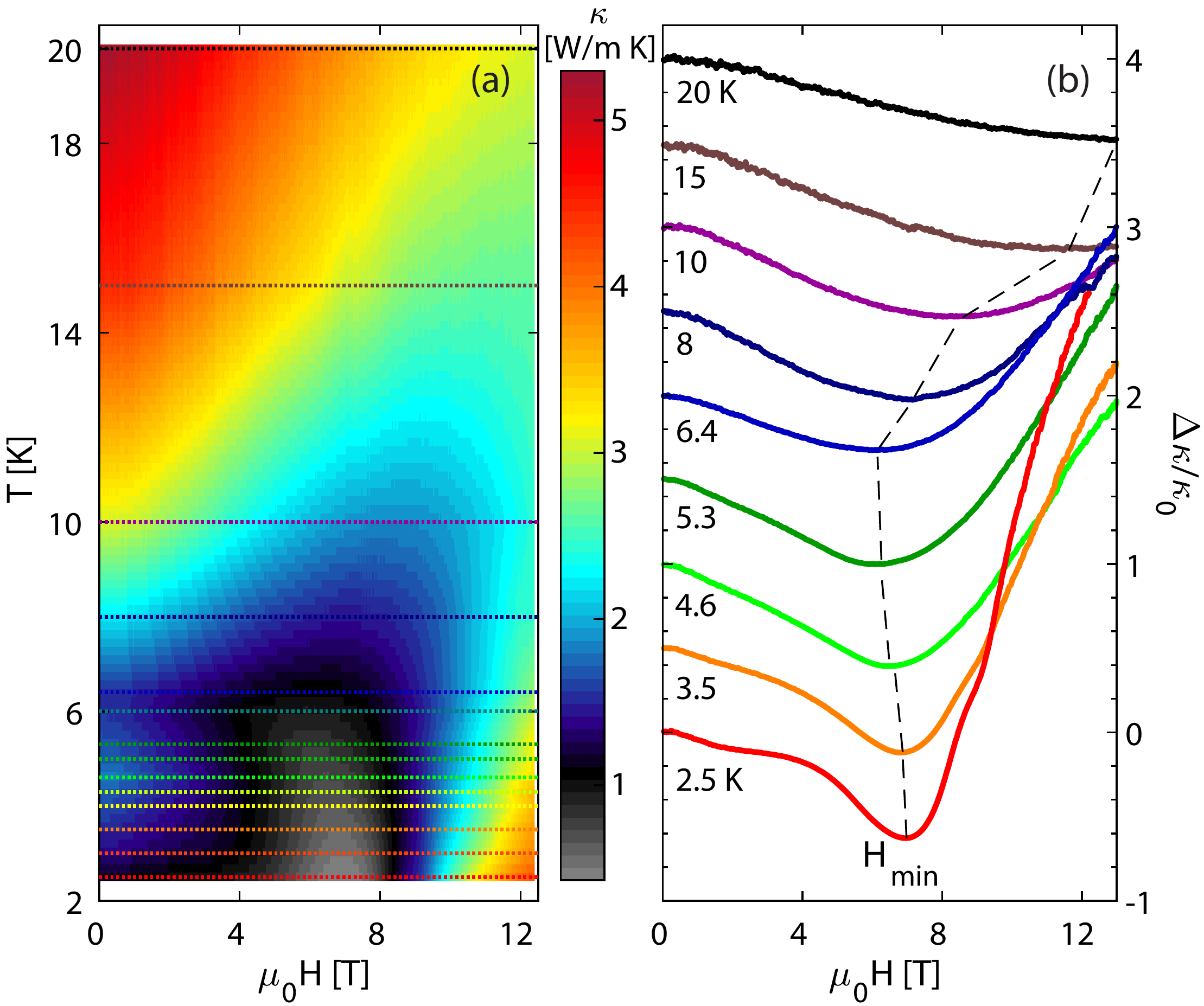}
\caption{(a) $\kappa(T,H)$ for Sample 1, represented by color contours. 
Horizontal lines correspond to field sweeps measuring $\kappa(H)$ at fixed 
$T$. (b) Relative thermal conductivity difference, $\Delta\kappa(H)/\kappa_0$, 
shown for the fixed values of $T$ highlighted in panel (a); curves are 
presented with constant offsets.}
\label{kappaB}
\end{center}
\efig

The non-monotonic evolution with $H$ and the strong high-field enhancement 
of $\ka$ are clearly evident in the isothermal $H$-dependence. Figure 
\ref{kappaB}(a) presents $\ka(H,T)$ for Sample 1 as a color contour map, 
showing the minimum region around $\Hmin$ and maxima at high $T$ or high $H$.   
The fractional change of $\ka(H)$, $\Delta\ka/\ka_0 = (\ka(H) - \ka_0)/\ka_0$, 
is shown in Fig.~\ref{kappaB}(b) for a range of $T$ values. $\ka(H)$ and 
$\Delta \ka(H)/\ka_0$ show an initial decrease, before turning over at 
$\Hmin$ and increasing rapidly. $\Hmin(T)$ remains around 7 T for $T < T_C$, 
but becomes rapidly larger as $T$ is increased beyond $T_C$, making the minima 
shallower until at $T = 20$ K $\Hmin$ is pushed outside our measurement range. 
Our measured value $\Hmin \simeq 7$ T for $T < 10$ K coincides with the 
critical field ($H_C$) for the field-induced phase transition observed in 
bulk magnetization \cite{Johnson2015} and specific-heat measurements 
\cite{Kubota2015}. Further, the magnetization in this field range is far from 
saturation \cite{Johnson2015,Kubota2015} and it is safe to conclude that the 
system is only weakly spin-polarized above $\Hmin$. 

In general, $\ka$ contains multiple terms whose effects can be difficult 
to separate. For $\alpha$-\rucl, the presence and location of $\Hmin$ are 
fundamental properties of the phase diagram and four further, distinctive 
features provide clues about the primary contributions to $\ka$. These 
are (i) the local minimum in $\ka(T)$ occurring at $T_C$ at small $H$ 
[Figs.~\ref{kappaT}(b) and \ref{kappaT}(c)], (ii) the slow decrease of 
$\ka(H)$ when $H$ increases from zero [Fig.~\ref{kappaB}(b)], (iii) 
the properties of the low-$T$ peak in $\ka(T)$ at $H > \Hmin$ 
[Figs.~\ref{kappaT}(b) and \ref{kappaT}(c)], and (iv) the suppression 
and shift of the intermediate-temperature contribution by the applied 
field [Fig.~\ref{kappaT}(a)].

Features (i) and (ii) can be explained within a conventional framework. 
The magnetic anisotropy of \rucl~results in a magnon gap of 1.7 meV 
\cite{Banerjee2015} in the ordered state, and thus no spin-wave 
contribution can be expected. In many systems, $\ka$ decreases rapidly below 
the Curie-Weiss temperature, $\Tcw$, due to the scattering of phonons by 
spin fluctuations, which reduces the phonon mean free path, $l_p$. Such 
spin-phonon scattering is thought to have a strong impact on the phonon 
contribution to heat conduction in SOC materials \cite{Steckel2016, 
Chernyshev2005}. In \rucl, this effect is visible below $\Tcw \approx 25$ K 
\cite{Sears2015} at ZF [Fig.~\ref{kappaT}(a)]. However, spin fluctuations 
are suppressed due to the onset of magnetic order, i.e.~below $T = T_C$. 
Thus the weak low-$T$, low-$H$ feature, whose vanishing causes the pronounced 
local minimum at $T_C$ in ZF (i), is caused by the enhancement of $\ka$ 
expected from the improved $l_p$. By the same token, the weak decrease of 
$\ka$ with $T$ for $H < \Hmin$ (ii) is a consequence of the applied field 
suppressing the magnetic order, and with it the improved $l_p$. 

\begin{figure}[t]
\begin{center}
\includegraphics[width=.90\linewidth]{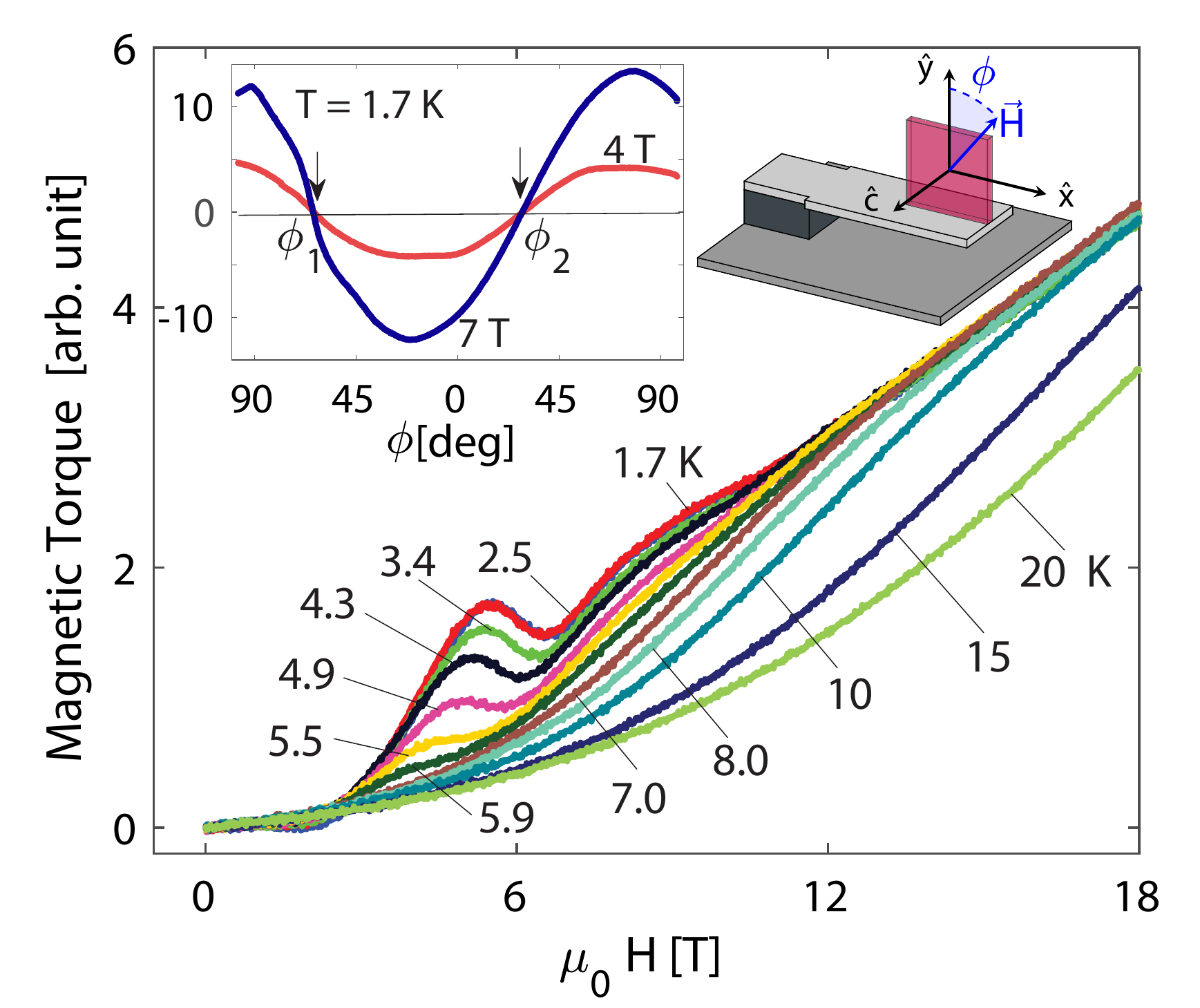}
\caption{In-plane torque response as a function of $H$, measured at selected 
values of $T$ with $\phi = - 69 \pm 2^{\circ}$. Right inset: measurement 
configuration. Left inset: $\phi$-dependence of $\tau$ at fields of 4 T 
and 7 T; $\phi_1 = - 64^{\circ}$ and $\phi_2 = 28^{\circ}$ exhibit 90$^{\circ}$ 
symmetry ($\phi = 0^{\circ}$ is chosen arbitrarily).}
\label{torque}
\end{center}
\efig

Before discussing features (iii) and (iv), for further perspective 
concerning the phases below and above $\Hmin$ we have performed magnetic 
torque measurements on our \rucl~single crystals. We rotate $H$ both within 
the $ab$-plane (Fig.~\ref{torque}, right inset) and out of it [discussed in 
Sec.~SII of the SM \cite{SuppInfo}]. The torque generated in the presence 
of a magnetization, $M$, is $\vec\tau = \mu_0 \mathbf{M} V \! \times \! 
\mathbf{H}$, with $\mu_0$ the permeability and $V$ the sample volume. The 
thermodynamic quantity $\tau$ is highly sensitive to magnetic anisotropy 
\cite{Okazaki2011,Kasahara2012}. Measurements performed on three crystals, 
of different shapes and sizes, all returned results very similar to those 
shown in Fig.~\ref{torque}. 

In the $ab$-plane, $\tau(\phi)$ displays the $90^{\circ}$ symmetry expected 
due to the monoclinic structure of $\alpha$-\rucl~\cite{Park2016, Cao2016} 
(Fig.~\ref{torque}, left inset). At two specific angles, $\phi_{1}$ and 
$\phi_{2}$, $\tau \rightarrow 0$ independent of the magnitude of $H$, and 
Fig.~\ref{torque} shows $\tau(H)$ measured near $\phi_{1}$ (results near 
$\phi_2$ are qualitatively similar). At low $T$, $\tau(H)$ with $H < \Hmin$ 
exhibits a strikingly non-monotonic form. This complexity ceases abruptly at 
$H > \Hmin$. At $T > T_C$, the sizes both of $\tau$ and of the anomaly drop 
significantly, indicating strongly that this behavior is due solely to the 
presence of magnetic order. Such anomalous $H$-dependence is not surprising 
in a Hamiltonian as anisotropic as the Kitaev-Heisenberg model, and a rich 
variety of complex field-induced ordering patterns, with corresponding 
off-diagonal components of the magnetic susceptibility tensor, has been 
suggested \cite{Janssen2016, Chern2016}. 

\begin{figure}[t]
\begin{center}
\includegraphics[width=.9\linewidth]{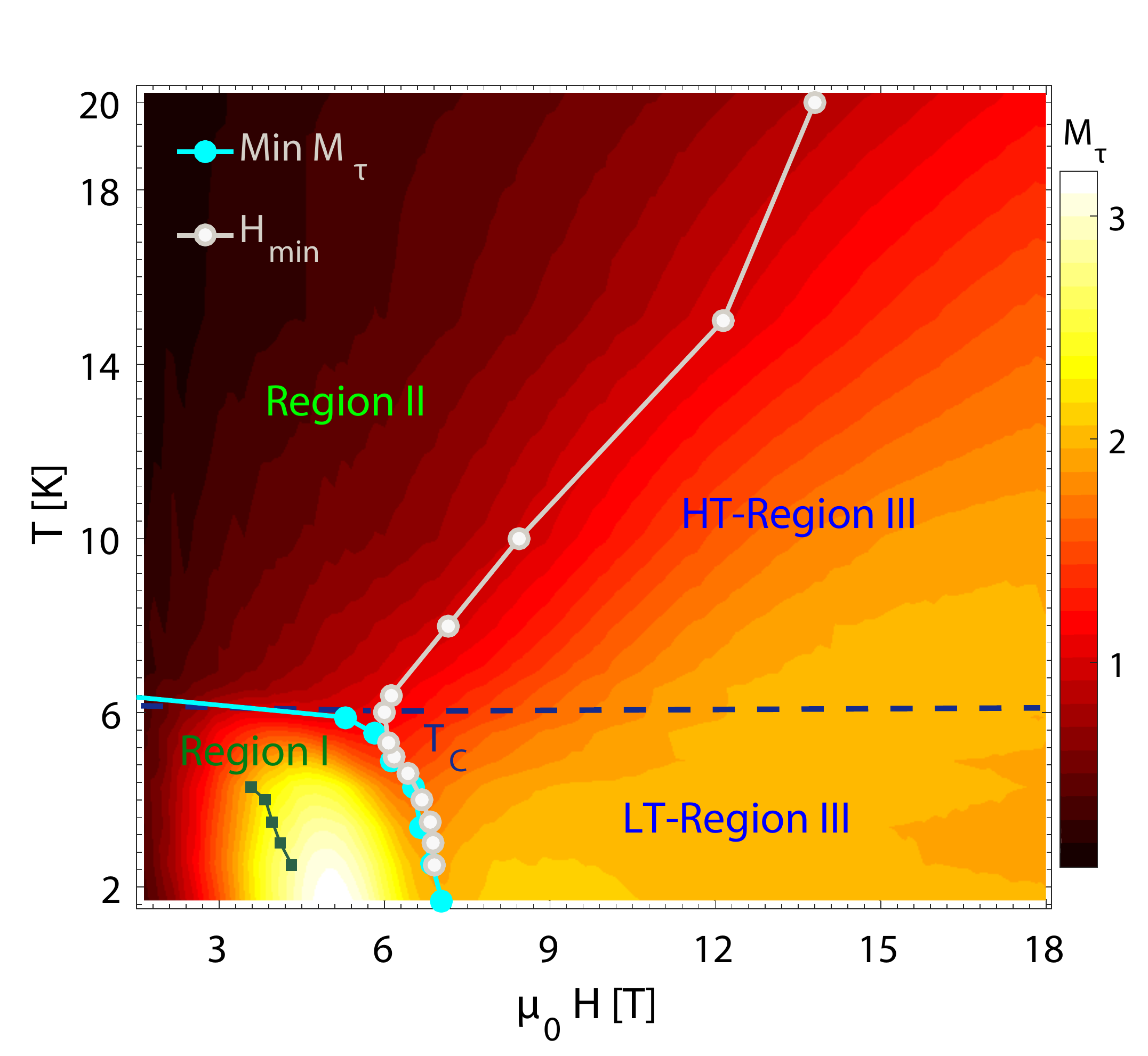}
\caption {(a) $(H,T)$ phase diagram of $\alpha$-\rucl~inferred from the 
magnitude of $\Mtau$ measured at $\phi \simeq - 69^{\circ}$ (color scale). 
White circles indicate $\Hmin$ as a function of $T$, cyan circles the 
position of local minima in $\Mtau(H)$, and green squares the locations of 
inflection points appearing in $\Delta\ka/\ka_0(H)$ [Fig.~\ref{kappaB}(b)].}
\label{phase}
\end{center}
\efig

We define the torque magnetization, $\Mtau = \tau/H$, which in certain 
geometries is closely related to the real magnetization (as discussed in 
Sec.~III of the SM \cite{SuppInfo}). Figure \ref{phase} shows $\Mtau(H,T)$ 
in the form of color contours \cite{note1}. The overlaid points showing the 
characteristic quantities $\Hmin(T)$ and the minima of $\Mtau$ divide this 
effective $H$-$T$ phase diagram naturally into three distinct regions. 
Region I, at $T < T_C$ and $H < \Hmin$, is where spontaneous magnetic order 
exists and is characterized by the strongly non-monotonic $\tau(H)$ and 
decreasing $\ka(H)$ ($d\ka/dH < 0$). Here also the inflection points of 
$\ka(H)$ at $H < \Hmin$ [Fig.~\ref{phase}(a)] coincide with the local maxima 
of $\tau$ (Fig.~\ref{torque}). In Region II, $d\ka/dH < 0$ while $T > T_C$. 
Region III is characterized by $d\ka/dH > 0$, but the derivative falls rapidly 
as $T$ crosses $T_C$, leading us to divide it into LT--Region III ($T < T_C$), 
where we observe the strongest enhancement of $\Delta\ka/\ka_0$, and 
HT--Region III ($T > T_C$), where $\Hmin$ moves rapidly to higher values.

Features (iii) and (iv) in $\ka$ are respectively the key properties of 
Regions LT--III and HT--III. Before invoking exotic physics, the conventional 
explanations should be exhausted. Modelling all the contributions of phonons 
and coherent spin excitations to $\ka$ is a complicated problem 
\cite{Ando1998,Sologubenko2000,Hofmann2001,Sologubenko2001,Sales2002,
Jin2003,Li2005,HessReview2007,Chen2011, Mohan2014}. The first complexity 
for $\alpha$-\rucl~is the quasi-2D structure, which would require a currently 
unavailable anisotropic 3D phonon fit. Conventional phonon thermal conductivity 
in a magnetic insulator is ascribed to four contributions, point defects, 
grain boundaries, Umklapp processes, and magnon-phonon resonant scattering 
\cite{Klemensbook,Callaway1959,Bergmanbook}. The last has been used 
successfully to describe the $H$-dependence of features observed in $\ka$ 
in several low-dimensional materials \cite{Hofmann2001,Wu2015,Jeon2016}. 
However, its effect is usually to generate a minimum at the resonance energy, 
causing a double-peak structure in $\ka(T)$ where only the lower peak has 
strong $H$-dependence \cite{Ando1998,Hofmann2001,Jeon2016}. Such behavior is 
qualitatively different from $\alpha$-\rucl, and in fact we are not aware of 
a mechanism for a strong field-induced enhancement of $\ka$, of the type we 
observe in Figs.~\ref{kappaT}(b) and \ref{kappaT}(c), other than a coherent 
spin excitation \cite{Sales2002, Jin2003, Li2005, Sun2009,Wang2010}. However, 
$\alpha$-\rucl~at $H > \Hmin$ has no magnetic order, as demonstrated by the 
absence of features in the susceptibility \cite{Sears2015} and specific heat 
\cite{Kubota2015}, and verified by neutron scattering \cite{Nagler2016} and 
nuclear magnetic resonance studies \cite{Baek2016}. Thus it has no conventional 
gapless mode, and this is why we conclude that feature (iii) must attributed 
to an unconventional excitation of the FISL. 

Turning to exotic solutions, the proximity of the zig-zag ordered state 
to a Kitaev QSL at ZF \cite{Rau2014} is strongly suggestive. However, we 
note that an applied field destroys many of the exotic properties of the 
Kitaev QSL \cite{Kitaev2006}, that the FISL is partially spin-polarized, 
that the Heisenberg terms have non-trivial effects \cite{Rau2014}, and that 
ideal honeycomb symmetry is broken in monoclinic $\alpha$-\rucl~\cite{Cao2016,
Kim2016}. It is nevertheless instructive to recall that the Kitaev model has 
an exact solution in terms of Majorana fermions, one of which is massless with 
linear dispersion \cite{Kitaev2006} while the others are massive. Spin 
excitations are Majorana pairs, or equivalently Majorana modes pinned to 
static fluxes \cite{Baskaran2007}, and are all massive. Although one recent 
numerical study reports a QSL state above a critical field in a model for 
\rucl~\cite{Yadav2016}, it was found to be gapped. The low- and medium-energy 
spin excitations of \rucl~at ZF have been mapped by recent inelastic neutron 
scattering studies \cite{Banerjee2015,Banerjee2016}. In addition to the gapped 
spin waves, one finds a broad continuum of excitations centered around 5 meV 
\cite{Banerjee2016}. At finite fields, no information is yet available, other 
than the $\ka$ signals we measure. In this context it is of crucial importance 
that the low-$T$ excitations contributing above $\Hmin$ are gapless and that 
their density of states, measured by $\ka$, increases linearly with $H$ 
[feature (iii)]. These are the properties of a cone-type dispersion and are 
thus the same behavior as the massless Kitaev QSL mode. 

Feature (iv) is the striking field-dependence of $\ka(T)$ for $T_C < T < 60$ K 
[Fig.~\ref{kappaT}(a)]. In this range $\ka$ should be dominated by phonons, 
whose contributions are $H$-independent. The presence of an incoherent 
medium-energy continuum of anisotropic spin excitations \cite{Banerjee2016} 
may cause a direct contribution to $\ka$ or a suppression due to spin-phonon 
scattering. Our results contain no evidence for direct contributions, as there 
is no abrupt change in $\ka$ at $\Hmin$ and the change in the high-$T$ peak 
position indicates energy shifts far beyond the scale of $H$. By contrast, 
our results contain several features characterizing a strong suppression of 
phonon contributions. First, $\ka$ at ZF cannot be fitted within the 
conventional framework, indicating that anomalous phonon scattering is 
significant even at $H = 0$. Second, the continuum affects the phonon 
contributions to $\ka$ over a broad range of $T$ [Fig.~\ref{kappaT}(a)], 
reflecting the broad energy range observed in Ref.~\cite{Banerjee2016}. 
Third, scattering becomes considerably more effective at a field of 14 T. 
Because the field scale for a significant reconstruction of the continuum 
should be the Kitaev energy, estimated as $K \approx 7$ meV \cite{Kim2015,
Banerjee2015}, it is clear that some rearrangement must take place at the 
field-induced transition to the FISL. However, the lack of abrupt changes in 
$\ka$ at $\Hmin$ indicates that only a small fraction of the continuum turns 
into the massless mode [feature (iii)], while the majority of its spectral 
weight remains in a broad continuum at finite energies; we comment here that 
thermal fluctuations may cause at least as strong a rearrangement of the 
continuum over the $T$ range of our experiment as field effects do. Thus 
from the evidence provided by $\ka(H,T)$, the FISL does appear to possess 
both the primary excitation features of the Kitaev QSL, namely a Dirac-type 
band and a finite-energy continuum. For this reason we refer to them as 
proximate Kitaev excitations (PKEs). 

To summarize the nature of heat conduction in $\alpha$-\rucl~in the context 
of Fig.~\ref{phase}, $\ka$ in Region I is controlled by low-$T$ phonon 
contributions, decreasing slowly as the system is driven towards the FISL 
because $l_p$ is reduced. A similar trend continues in Region II, where 
the increasing thermal population of phonons, as well as thermally excited 
paramagnons, contribute to $\ka$. In LT--Region III, the rapid increase of 
$\ka$ with $H$ reflects the presence of the massless PKE. In HT--Region III, 
the minimum of $\ka(H)$ moves to high fields (Fig.~\ref{phase}) and there 
are contributions to $\ka$ both from phonons and from the massive PKEs, where 
the primary effect of the latter is the systematic suppression of the former 
with increasing $H$, which drives up the crossover field [$\Hmin(T)$] from 
Region II.

To conclude, we have investigated the highly non-monotonic thermal conductivity 
and the torque magnetization response of the 2D honeycomb-lattice material 
\rucl. We infer a field-induced phase transition to a state, the FISL, of no 
magnetic order and no simple spin polarization. The low-energy excitations of 
this spin-disordered ground state cause a dramatic enhancement of $\ka$ at low 
temperatures, while its gapped excitations suppress the phonon contribution 
at higher temperatures, and do so more effectively at higher fields. Although 
our results neither prove nor disprove that the FISL is closely related to 
the Kitaev QSL state, they set strong constraints on the nature of its 
excitations and thus of its theoretical description.


\vspace{0.1in}

\noindent{\bf {Acknowledgment}} We thank A. Chernyshev and M. Hermele for 
helpful discussions. This work was supported by the US DOE, Basic Energy 
Sciences, Materials Sciences and Engineering Division, under Award Number 
DE-SC0006888. Torque magnetometry was performed at the National High Magnetic 
Field Laboratory, which is supported by National Science Foundation Cooperative 
Agreement No.~DMR-1157490 and the State of Florida.


\newpage

\setcounter{page}{1}
\setcounter{equation}{0}
\setcounter{figure}{0}
\renewcommand{\theequation}{S\arabic{equation}}
\renewcommand{\thefigure}{S\arabic{figure}}

\onecolumngrid

\vskip8mm

\section*{\Large{Supplemental Material}}

\centerline{\bf {Anomalous thermal conductivity and magnetic 
torque response in the honeycomb magnet $\alpha$-RuCl$_3$ }}
  
\vskip3mm

\centerline{Ian A. Leahy, Christopher A. Pocs, Peter E. Siegfried, David Graf, 
S.-H. Do, Kwang-Yong Choi,} 

\centerline{B. Normand, and Minhyea Lee}

\vskip6mm

\twocolumngrid

\subsection{SI. Sample Synthesis and Sample-Dependence}

\subsubsection{Sample Synthesis}

The $\alpha$-\rucl~single crystals used in this study were grown by a 
vacuum sublimation method.  A commercial \rucl~powder (Alfa-Aesar), which 
consists of $\beta$-\rucl~with a small fraction of RuO$_2$ and RuOCl$_2$, was 
ground thoroughly and dehydrated in an evacuated quartz ampoule. When the 
sealed ampoule is heated at 600$^{\circ}$C for one day, the $\beta$-phase of 
RuCl$_3$ undergoes a transition to the $\alpha$-phase, while RuOCl$_2$ 
decomposes to RuO$_2$, $\alpha$-RuCl$_3$, and Cl$_2$ gas. The remaining RuO$_2$ 
in the mixture intervenes in the $\alpha$-RuCl$_3$ powder phase. During the 
course of crystallization, single crystals of RuO$_2$ grow separately from 
the $\alpha$-RuCl$_3$ crystals, allowing their removal by mechanical treatment, 
and thus no additional purification was required. The ampoule was placed in a 
temperature-gradient furnace for vapor transport. The mixture was heated to 
1080$^{\circ}$C, then cooled to 650$^{\circ}$C at a rate of $-2^{\circ}$C per hour. 
We obtain pieces of $\alpha$-RuCl$_3$ single crystals at the end of the 
ampoule. Single-crystal diffraction and EDX (Electron Dispersive X-ray) 
measurements confirm that these $\alpha$-RuCl$_3$ crystals are single-phase.

\subsubsection{Sample-dependence in $\ka$ and $\tau$ measurements}

We performed experiments on three different $\alpha$-\rucl~crystals, which 
we have labelled in time order. Figure 1 of the main text shows our results 
for $\ka(T)$ for Samples 1, 2, and 3 at selected field values. We reiterate 
that the primary features of $\ka(T)$, namely $T_C$, $H_{\rm min}$, the peak 
positions, and the trends in peak heights, are little changed between samples. 
However, it is clear that quantitative details of our measurements do differ 
between samples, particularly the heights of the peak features for $H$ both 
below and above $H_{\rm min}$ in Fig.~1(a) and the widths and centers of the 
low-$T$ ($T < 10$ K) peaks of Samples 1 and 2, which are compared in 
Figs.~1(b) and 1(c). 

For completeness, we show in Fig.~\ref{figs1} the $H$-dependence of $\ka$ for 
Samples 1 and 2; $\ka(H)$ of Sample 1 is shown in scaled form in Fig.~2(b) of 
the main text, and here we compare it directly with Sample 2. The relative 
changes in $\ka$ induced by the applied field are discernibly smaller for 
Sample 2 than for Sample 1. However, as above the overall characteristics of 
the $H$-dependence, namely the pronounced minima at $H = H_{\rm min} \approx 7$ 
T for $T < T_C$, the rapid increase at $H > \Hmin$, and the movement of 
$H_{\rm min}$ to higher fields with increasing $T$, are consistent across the 
different samples. We state here that we do not have comparable measurements 
for Sample 3. Regrettably, because this appears from Fig.~1(a) of the main 
text to be our highest-quality crystal, we suffered a sample breakage after 
our first measurements with Sample 3, and this is why we have no data 
comparable either with Fig.~S1 or with Figs.~1(b) and 1(c). 

\bfig[t]
\begin{center}
\includegraphics[width=0.9\linewidth]{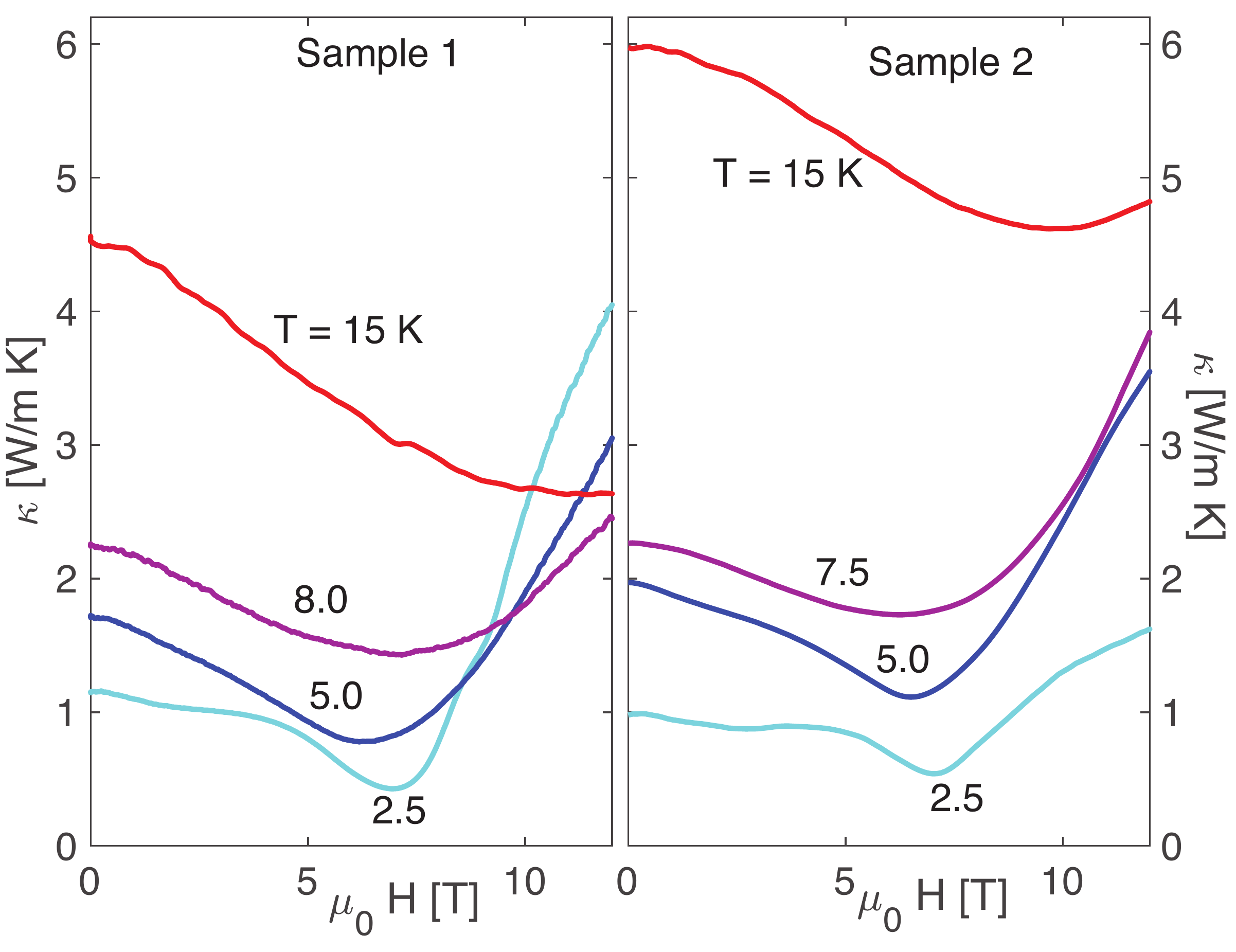}
\caption {Thermal conductivity, $\kappa(H)$, of (a) Sample 1 and (b) Sample 2, 
shown at selected values of $T$.}
\label{figs1}
\end{center}
\efig

By contrast, the samples used in our magnetic torque measurements showed 
very consistent results. These measurements are made with much smaller 
crystals and were repeated three separate times in the course of 9 months. 
The size of the samples used for torque measurements is typically 100--200 
$\mu$m, which is more than 10 times smaller in linear dimension than those 
used for the measurement of $\ka$. 

\subsubsection{Sources of sample-dependence}

The variations we observe in the low-temperature thermal conductivity 
indicate a non-negligible role for different extrinsic factors, including 
impurities, defects, stacking faults, and magnetic domains. In one 
well-documented example, the authors of Ref.~\cite{Cao2016} showed that 
limited physical working of $\alpha$-\rucl~can act to alter the $c$-axis 
stacking arrangement of the honeycomb layers, which has a direct effect on 
the magnetic transition temperature. Although our pristine crystals were 
nearly free from stacking faults, and from the formation of associated 
domains, it is quite plausible that different mechanical forces were 
applied on the different samples during their preparation for the $\ka$ 
measurements. However, we were not able to find any correlations between 
the sample-dependent thermal conductivity and other basic characterization 
results.

\begin{figure}[t]
\begin{center}
\includegraphics[width=0.95\linewidth]{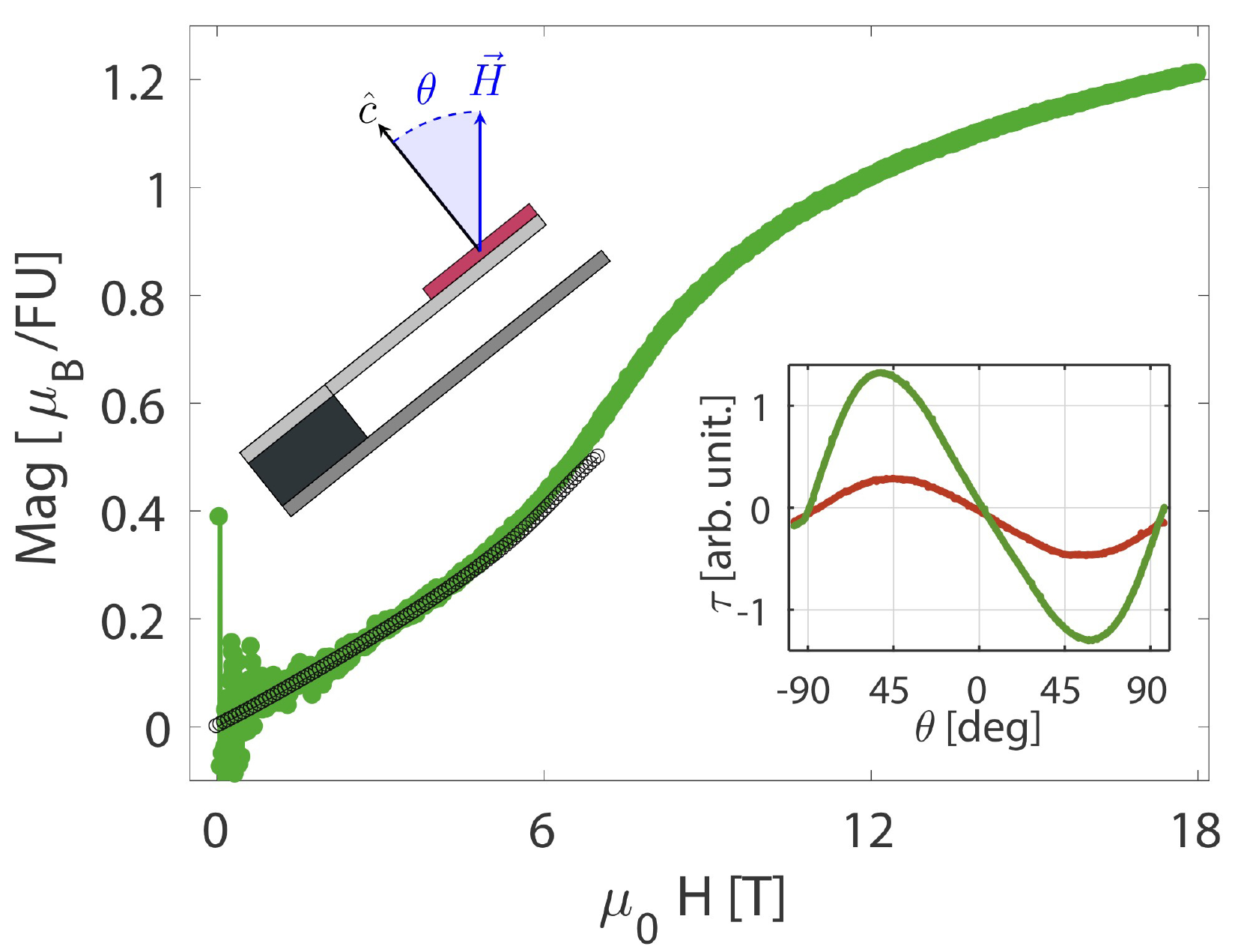}
\caption{Torque magnetization, $M_{\tau} = \tau/H$, measured at $T = 1.5$ K 
in the geometry shown in the left inset with $\theta \simeq 87^{ \circ}$. 
Results for $M_{\tau}$ are scaled to match magnetization data measured up to 
$\mu_0 H = 7$ T with a SQUID magnetometer, which are shown as black circles. 
Right inset: $\tau$ as a function of $\theta$ at fixed $H$. The sample 
magnetization is expected to remain largely within the $ab$ plane over 
most of the range of $\theta$.}
\label{figs3}
\end{center}
\efig

Turning to the other extrinsic sources for quantitative differences, although 
our measurements were performed over an extended period we did not observe 
significant effects from sample aging; while we tried to minimize extended 
exposure to air where possible, these effects would include air-sensitivity 
and surface degradation. As a result, we did not record much sample variation 
of the dc susceptibility. However, we do infer a significant difference 
between samples in the density of magnetic domains and defects, two quantities 
to which the thermal conductivity is considerably more sensitive than are 
bulk probes (dc susceptibility and also specific heat). In addition to the 
$c$-axis stacking, other causes of domain formation include the threefold 
degeneracy of the zig-zag ordered structure and twinning in the monoclinic 
crystal structure. The smaller sample-dependence observed in the magnetic 
torque suggests a relatively smaller sensitivity of this quantity to the 
presence of domains. 

\subsection{SII. Magnetic torque measured by out-of-plane field rotation}

The strong magnetic anisotropy between the $ab$-plane and the $c$-axis 
makes it possible to recover a reliable estimate of the in-plane 
magnetization from measurements of the torque. Because $\vec\tau = \mu_0V 
\vec M \times \vec H$, the magnitude of the torque is given by $MH \sin 
\theta'$, where $\theta'$ is the angle between $\vec M $ and $\vec H$. In 
the geometry shown in the left inset of Fig.~\ref{figs3}, the easy-plane 
magnetic anisotropy means that $\theta' \rightarrow 0$ when $\theta 
\rightarrow 90^{\circ}$ and therefore $|\vec\tau| \rightarrow 0$. Similarly, 
when $\theta \rightarrow 0$, i.e.~$H$ is aligned to the hard axis, the 
component of $\vec M$ aligned to the field is almost zero and the measured 
$\tau \rightarrow 0$. This is illustrated in the $\theta$-dependence of 
$\tau$ at fixed $H$, shown in the right inset of Fig.~\ref{figs3}. The 
$H$-dependence of $\Mtau$ deduced from torque measurements performed at 
$\theta = 87^{ \circ} \pm 2^{ \circ}$, where $2^{ \circ}$ is the experimental 
uncertainty in the out-of-plane angle, gives a good approximation to the 
averaged in-plane magnetization. The magnitude of $\Mtau$ is calibrated 
by matching to the magnetization measured by a SQUID magnetometer. Our 
results (Fig.~\ref{figs3}) are fully consistent with the bulk magnetization 
reported in Ref.~\cite{Kubota2015}. 

\bfig[t]
\begin{center}
\includegraphics[width=0.95\linewidth]{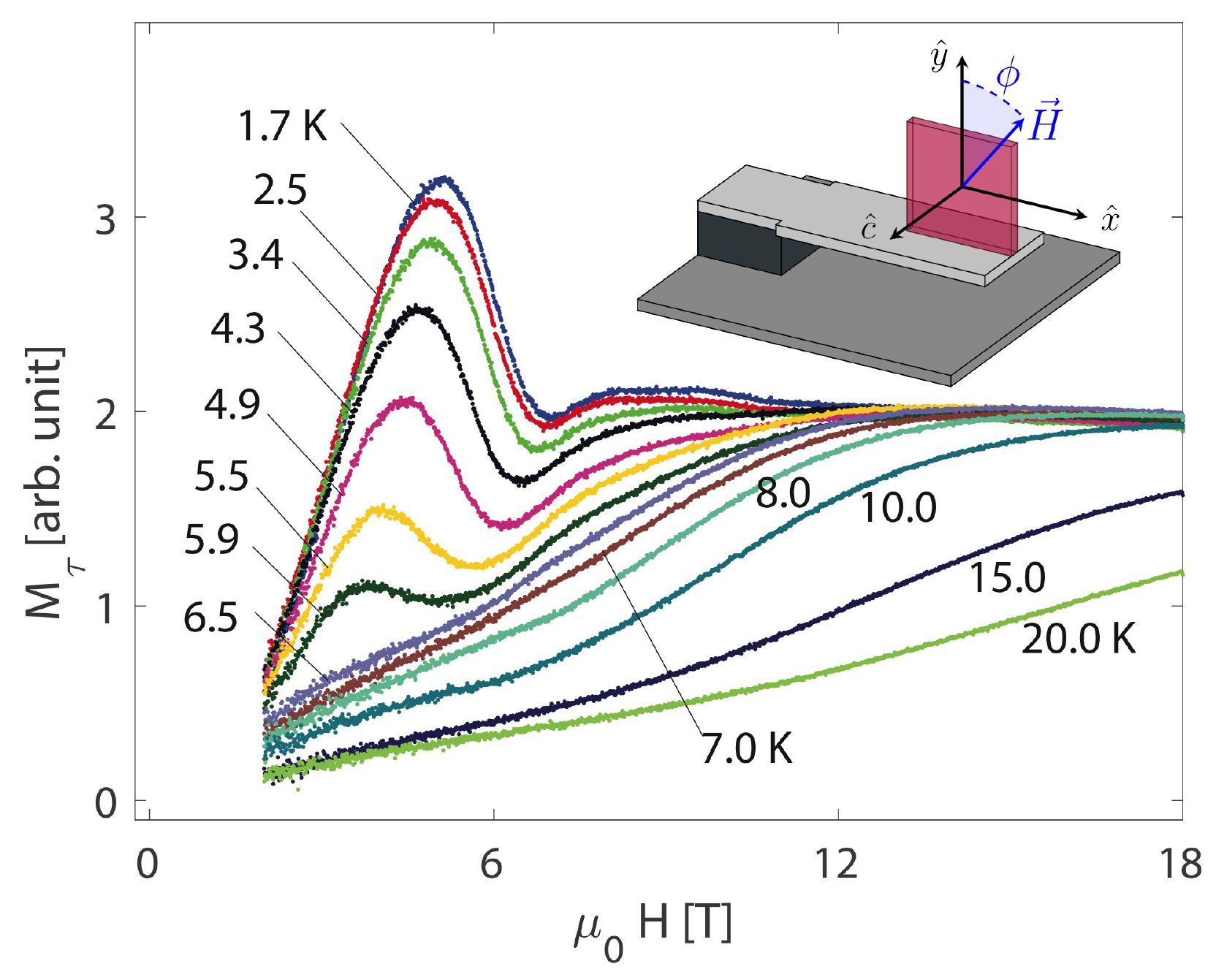}
\caption {Torque magnetization, $\Mtau(H)$, measured with $H$ applied in 
the $ab$-plane. Inset: schematic representation of measurement geometry. 
Field sweeps were performed at $\phi = -69^{\circ} \pm 2^{\circ}$, which is 
approximately 5$^{\circ}$ away from the angle where $\tau \rightarrow 0$.}
\label{figs2}
\end{center}
\efig

\subsection{SIII. Torque magnetization measured by in-plane field rotation}

Figure \ref{figs2} shows the torque magnetization, $\Mtau = \tau(H)/H$, 
obtained by $\phi$ rotation (inset). These data display directly the 
information shown by the color contours in Fig.~4 of the main text and 
the corresponding data for $\tau(H)$ are shown in Fig.~3. We reiterate 
that non-monotonic $H$-dependence appears only for $H < \Hmin \simeq 7$ T, 
beyond which the data converge to a constant value of $\Mtau$ if $T < 10$ 
K. Further increase of $T$ causes the magnitude of $\Mtau$ to drop 
significantly. Because the out-of-plane magnetic anisotropy is poorly 
understood, unlike the case of $\theta$ rotation (Sec.~SII) the decomposition 
of $\Mtau$ from $\phi$ rotation into two or more orthogonal components is 
not trivial. The torque data used here were measured at $\phi \simeq - 
69^{ \circ}$ (inset, Fig.~3, main text), a position in which $M$ is very well 
aligned with $H$. 

\end{document}